\documentclass[preprint2,6pt]{emulateapj}%
\usepackage{graphicx}
\usepackage{amsmath}
\usepackage{amsfonts}
\usepackage{amssymb}%
\providecommand{\U}[1]{\protect\rule{.1in}{.1in}}

\newcommand{\cjaa}{{\it Chinese~J.~Astron.~Astrophys.}}

\newcommand{\beq}{\begin{equation}}
\newcommand{\eeq}{\end{equation}}
\newcommand{\ba}{\begin{array}}
\newcommand{\ea}{\end{array}}

\begin{document}

\title{Dynamical Properties of Internal Shocks Revisited}
\shorttitle{Dynamics of Internal Shocks}
\author{Asaf Pe'er \altaffilmark{1}, Killian Long \altaffilmark{1}, and Piergiorgio Casella \altaffilmark{2} }
\shortauthors{Pe\'er, Long \& Casella}

\altaffiltext{1}{Physics Department, University College Cork, Cork, Ireland}
\altaffiltext{2}{INAF, Osservatorio Astronomico di Roma, Via Frascati 33, I-00078 Monteporzio Catone, Italy}

\begin{abstract}
Internal shocks between propagating plasma shells, originally ejected
at different times with different velocities are believed to play a
major role in dissipating the kinetic energy, thereby explaining the
observed lightcurve and spectra in a large range of transient objects.
Even if initially the colliding plasmas are cold, following the first
collision the plasma shells are substantially heated, implying that in
a scenario of multiple collisions, most collisions take place between
plasmas of non-zero temperatures. Here, we calculate the dynamical
properties of plasmas resulting from a collision between arbitrarily hot
plasma shells, moving at arbitrary speeds. We provide simple
analytical expressions valid for both the ultra-relativistic and
Newtonian velocities, for both hot and cold plasmas. We derive the
minimum criteria required for the formation of the two-shock wave
system, and show that in the relativistic limit, the minimum Lorentz
factor is proportional to the square root of the ratio of the initial
plasmas enthalpies. We provide basic scaling laws of synchrotron
emission from both the forward and reverse shock waves, and show how
these can be used to deduce the properties of the colliding
shells. Finally, we discuss the implications of these results in the
study of several astronomical transients, such as x-ray binaries,
radio loud quasars and gamma-ray bursts.

\end{abstract}
\maketitle

\section{Introduction}

Lightcurves of many astronomical transients that are characterized by
strong outflows (jets) show substantial variability, observed on
timescales as fast as millisecond and possibly even faster. Several
examples include blazars \citep{Marscher80}, gamma-ray bursts
\citep[GRBs;][]{Norris+96} and x-ray binaries
\citep[XRBs;][]{Fender01}. A leading model proposed to explain these
variable lightcurves is the internal shocks model. The basic idea is
that variability within the inner engine results in fluctuations in
the ejection of plasmas. Thus, the ejected material propagates as a
collection of ``plasma shells''. Each individual shell is accelerated
and then propagates at some terminal velocity which is independent on
the terminal velocities of the other plasma shells. At a second stage,
shells that were ejected at later times but with faster speeds, catch
up with the slower shells ahead. The collision between the plasma
shells results in the formation of two shock waves (forward and
reverse) that propagate, respectively, into the slow and fast
shells. These shock waves dissipate part of the shells' kinetic
energy, which is then radiated away. Following the collision, the
colliding shells are assumed to merge and continue propagating
together (i.e., the collision is considered as a plastic collision),
and are therefore subject to a subsequent collision with a third
incoming, faster shell. This scenario of multiple collisions therefore
results in the observed variable lightcurve.
 
Such models were proposed to explain the knots in active galactic
nuclei [AGN] jets \citep{Rees78}. They have been in wide use since the
1990's in explaining the rapid variability observed during the prompt
phase of many GRBs \citep[e.g.,][]{RM94, Fenimore+96, SP97,
  Kobayashi+97, DM98, PSM99, RF00, GSW01, MRRZ02, NP02, KMY04,
  Canto+13}, as well as in blazars \citep{SBR94, Ghis99, Spada+01,
  BD10, MA10}.  In recent years, similar models were applied in the
study of variable emission from XRBs \citep{KSS00, Miller-Jones+05,
  Jamil+10, Malzac13, Malzac14, Drappeau+15} as well as tidal
disruption events \citep{WC12}. See \citet{Peer14} for a review on the
similarities between these objects.  Indeed, the hydrodynamical
properties of the shock waves as well as the colliding shells had long
been investigated in the non-relativistic as well as in the
relativistic regimes \citep{BM76, SP95}.

Despite its considerable popularity, it should be stressed that it is
still unclear today whether internal shocks by themselves are the
leading mechanism that produces the observed signal in these
objects. In the context of GRBs for example, there are two main
drawbacks of this model. Firstly, the relatively low efficiency in
energy conversion, as only the differential kinetic energy can be
dissipated. Several authors found that the typical efficiency of
energy conversion that can be expected in a multiple shell collisions
is only a few - few tens of \% \citep{Kobayashi+97, DM98, Kumar00,
  GSW01, FW01, Ioka+06}. This result, however, depends on the velocity
distribution of the ejected shells, and can become substantially
higher under the appropriate conditions \citep{Beloborodov00,
  KS01}. The second drawback (in the context of GRBs) is accumulating
evidence in recent years that a thermal component may play an
important role in explaining at least part of the observed spectra in
a significant minority of GRBs \citep{Ryde04, Ryde05, Peer08, RP09,
  Lazzati+09, Guiriec+11, Ryde+11, Axelsson+12}. As this component
originates from the photosphere, a dominant thermal component implies
that a substantial energy dissipation at larger radii may be
unnecessary. Nonetheless, in most cases in which a thermal component
is observed, it is accompanied by an addition non-thermal part
\citep[e.g.,][and references therein]{Peer15}. 

While the validity of the internal shocks model as a leading energy
dissipation mechanism is uncertain, the main alternative dissipation
models, namely magnetic reconnection \citep{Coroniti90, Usov92,
  Thompson94, Drenkhahn02, LK01, DS02} suffers an even higher degree
of uncertainty. For example, as the rate of reconnection depends on
the magneto-hydrodynamic (MHD) turbulence, it is difficult to be
assessed from first principles without detailed specification of the
environment. It is therefore of no surprise that no consensus on the
origin of dissipation had been achieved to date.  An in depth
discussion on the current observational status in GRBs and its
implications appears in several recent reviews \citep[e.g., ][and
  references therein]{Piran04, WB06, Meszaros06, FM06, Zhang07, GRF09,
  MR14, Zhang14, KZ15}.  

Given these uncertainties, a more in depth study is needed on the
underlying physics of the different models.  Indeed, within the
framework of the internal shocks model, one thing in common to nearly
all studies carried so far is that a detailed description of the
shocked plasma conditions were calculated based on the assumption that
the colliding plasmas are initially cold \citep{SP95}. While the
expanding plasma shells lose their energy adiabatically during the
expansion, the shock waves formed in each collision substantially heat
the plasma. Thus, even if initially the plasmas are cold, in a
scenario of multiple collisions, in general the colliding plasmas are
not expected to be cold.  While this fact was considered by several
authors in calculating the overall efficiency of energy conversion
\citep{PSM99, KP00, Beloborodov00, Spada+01, GSW01, KMY04, Jamil+10,
  Malzac14}, so far no detailed description of the shocked plasma
properties was calculated in the general scenario of arbitrary hot
plasmas colliding at arbitrary velocities.

Such calculation is of particular importance for two reasons. First,
when arbitrary hot plasmas collide, the conditions for the formation
of the two shock wave system are not always fulfilled. As a result,
the amount of energy dissipated in such a collision can be
substantially smaller than if shock waves are formed. Second, even if
shock waves are formed, the expected spectra depends on the energy
density and energy per particle in the shocked region, which are in
general different than in the cold plasma collision scenario
\citep{ZM02a}. Thus, in order to provide accurate calculations of the
expected lightcurve and spectra, the conditions at the shocked plasma
needs to be determined.

A scenario of cold shell interacting relativistically with a hot ($e
\gg n m_p c^2$, where $e$ and $n$ are the energy and number densities,
$m_p$ is the proton mass and $c$ is the speed of light) expanding
shell that was slowed down by interacting with the ambient medium was
considered by \citet{KP00}. A relative Lorentz factor between the
colliding shells of 1.25 was found to represent well the scenario
considered in that work. The results indeed indicate that the plasma
properties following the collision deviates from the plasma properties
expected in the cold-cold shells collisions.

In this work, we calculate the properties of the shocked plasma
following the collision of two arbitrarily hot plasma shells. We consider a
simple 1-d model which enables us to provide simple analytic
estimations of the thermodynamic properties of the shocked plasmas in
the various regimes. As we show below, one needs to discriminate not
only between the relativistic and non-relativistic scenarios, but the
analytical solutions also depend on the energy densities of the
plasmas. We thus discriminate between ``cold'', ``cool'' and ``hot''
plasmas (see definitions in \S\ref{sec:2} below). We derive the
minimum criteria for the formation of such shocks in the different
scenarios, as well as the properties (velocity, energy density and
energy per particle) of the shocked plasma. In \S\ref{sec:3} we
provide a full numerical solution, which can be used for arbitrary
plasma properties, and further serves to validate and demonstrate the
analytical approximations in the different regimes. In \S\ref{sec:4}
we discuss observational consequences of the model, and in particular
we show that the efficiency in energy conversion is different than
previous claims, due to the need to include a pressure term. We
further demonstrate how the properties of the synchrotron emission can
be used to probe the properties of the colliding shells, before
summarizing in \S\ref{sec:5}.

\section{Basic Setup}
\label{sec:2}

We consider a slab of (non-magnetized) plasma shell (``slow shell''),
that propagates at some arbitrary speed $\beta_1 = v_1 / c$
(corresponding Lorentz factor $\Gamma_1$) in the lab frame. A second
plasma shell (``fast shell''), that propagates at velocity $\beta_4 >
\beta_1$ collides with the slow shell. At sufficiently high $\beta_4$,
two shock waves are formed as a result of the collision: a forward
shock propagating into the slow shell, and a reverse shock propagating
into the fast shell. A contact discontinuity separates the shocked
slow shell material from the shocked fast shell material.

Following the collision, there are four different regimes: (1) the
slow shell, (2) the shocked slow shell, (3) the shocked fast shell,
and (4) the fast shell material. The velocities of the four regimes
are $\beta_i$, $(i=1..4)$ and the corresponding Lorentz factors
$\Gamma_i = (1- \beta_i)^{-1/2}$. The thermodynamical quantities:
$n_i$, $e_i$, $p_i$ and $\omega_i = e_i + p_i$ (number density,
internal energy density, pressure and enthalpy, respectively) are
measured in each of the fluid's (shells) rest frame.  We further denote
the speed of the forward and reverse shock waves in the lab frame by
$\beta_{fs},~\beta_{rs}$, respectively (corresponding Lorentz factors
$\Gamma_{fs},~\Gamma_{rs}$). 

The system considered therefore contains a total of 18 free parameters
($\beta_i, n_i, e_i, p_i$ [$i=1..4$], $\beta_{fs},~\beta_{rs})$. The
shock jump conditions, namely conservations of particle, energy and momentum
flux densities at each shock wave provide six equations. Two more
equations are provided by equating the pressures and velocities along
the contact discontinuity. Additional four equations of state, of the
form $p_i = (\hat \gamma_i -1) (e_i - n_i)$, where $\hat \gamma_i$ is
the adiabatic index in region $i$, complete a total of 12 equations
connecting the velocities and thermodynamic quantities in all four
regimes. Thus, by specifying a total of six boundary conditions, namely
the initial velocities, number and energy densities in the slow and
fast plasma shells ($\beta_1, \beta_4, n_1, n_4, e_1, e_4$), the
velocities and thermodynamic properties of all four regions of the
system are fully determined.\footnote{We assume that the conditions
  are homogeneous within each regime. This approximation is reasonable
  as long as the initial colliding shells are not too wide. It can
  easily be justified in the shocked regions, which are sub-sonic.}

In this and the following sections we provide a complete solution in
the planar case. Clearly, in the most general scenario the values of
all 18 parameters can only be determined numerically. However, as we
show here, simple analytical solutions exist in the limiting cases of
relativistic ($\Gamma_4 \gg \Gamma_1$) as well as Newtonian ($\beta_4,
\beta_1 \ll 1$) plasma shells velocities. In this section we first
derive the analytical solutions, before providing a few examples of
the full numerical solution in \S\ref{sec:3} below.

\subsection{Analytical Solution: Basic Equations}
\label{sec:2.1}

For simplicity, we assume in the calculations below that the slow
plasma is at rest ($\beta_1 = 0,~\Gamma_1 = 1$). This can be viewed
both as a specific case, but equally as conducting the calculations in
the rest frame of the slow plasma shell. Towards the end of the next
section, we transform the derived results to the lab frame, in which
$\Gamma_1 \geq 1$ is arbitrary.  Thus, in the calculations below,
$\Gamma_2$ is the Lorentz factor of the shocked slow plasma (region (2))
in the rest frame of region (1), etc.

The forward shock jump conditions follow from the continuity of energy
($T^{01} = \omega \Gamma^2 \beta$), momentum ($T^{11} = \omega
\Gamma^2 \beta^2 + p$) and particle ($n \Gamma \beta$) flux densities
in the shock frame. Here, $T^{\mu \nu} = \omega u^\mu u^\nu + p
\eta^{\mu \nu}$ is the stress- energy tensor, $u^\mu$ is the
4-velocity and $\eta^{\mu \nu}$ is the metric tensor.  In their most
general form, the forward shock jump conditions can be put in the form
\citep{BM76, Wiersma07}
\beq
{e_2 \over n_2} = \Gamma_2 {\omega_1 \over n_1} - {p_1 \over n_2},
\label{eq:1}
\eeq
\beq
p_2 - p_1 = {(\Gamma_2 \beta_2)^2 n_2 \omega_1 \over n_2 - \Gamma_2 n_1}
\label{eq:2}
\eeq
Similarly, the reverse shock jump conditions are written as 
\beq
{e_3 \over n_3} = \bar \Gamma_3 {\omega_4 \over n_4} - {p_4 \over n_3},
\label{eq:3}
\eeq
\beq
p_3 - p_4 = {(\bar \Gamma_3 \bar \beta_3)^2 n_3 \omega_4 \over n_3 - \bar \Gamma_3 n_4}.
\label{eq:4}
\eeq
Here, $\bar \Gamma_3 = \Gamma_3 \Gamma_4 (1 - \beta_3 \beta_4)$ is the
Lorentz factor of the shocked material in region (3) relative to the
unshocked fast shell in region (4), and $\bar \beta_3 = (1-{\bar
  \Gamma_3}^{-2})^{1/2}$ is the corresponding velocity.

\subsection{Relativistic collision}
\label{sec:2.2}

In the ultra-relativistic case, we consider the scenario where
$\Gamma_4 \gg \Gamma_2 = \Gamma_3 \gg 1$. Under this assumption,
\beq
\bar \Gamma_3 \simeq {1 \over 2} \left( {\Gamma_4 \over \Gamma_3} +
     {\Gamma_3 \over \Gamma_4} \right) \approx {\Gamma_4
       \over 2 \Gamma_2}.
\label{eq:5}
\eeq
Since it is always true that $n_2 \geq n_1$ and $n_3 \geq n_4$ (and
clearly $\omega_1 \geq p_1,~ \omega_4 \geq p_4$), it is safe to
neglect the second terms in the right hand sides of equations
\ref{eq:1} and \ref{eq:3}. Using the equations of state for regions
(2) and (3), with the help of the modified Equations \ref{eq:1} and
\ref{eq:3}, the pressures in regions (2) and (3) can be written as
\beq
\ba{lcl}
p_2 & = & (\hat \gamma_2 -1) n_2 \left( {e_2 \over n_2} -1 \right)
\simeq (\hat \gamma_2 -1) n_2 \Gamma_2 { \omega_1 \over n_1}; \\
p_3 & \simeq & (\hat \gamma_3 -1) \omega_4 \bar \Gamma_3 { n_3 \over
  n_4}.
\ea
\label{eq:6}
\eeq
Using these results in Equations \ref{eq:2}, \ref{eq:4}, and
neglecting $p_1 \ll p_2$ as well as $p_4 \ll p_3$ (which is correct in
the ultra-relativistic limit), one obtains
\beq
{n_2 \over n_1} = {\hat \gamma_2 \over \hat \gamma_2 -1} \Gamma_2;
~~~{n_3 \over n_4} = {\hat \gamma_3 \over \hat \gamma_3 -1} \bar
\Gamma_3.
\label{eq:7}
\eeq
In order to proceed, we use the fact that in the relativistic limit,
the adiabatic indices are $\hat \gamma_2 = \hat \gamma_3 = 4/3$. Using
these results, as well as the approximation derived in Equation
\ref{eq:5} in Equation \ref{eq:6}, the requirement $p_2 = p_3$ leads
to
\beq
\Gamma_2 = \Gamma_3 \simeq \sqrt{\Gamma_4 \over 2} \left({\omega_4
  \over \omega_1}\right)^{1/4}, ~~~ \bar \Gamma_3 \simeq
\sqrt{\Gamma_4 \over 2} \left({\omega_1 \over \omega_4}\right)^{1/4}.
\label{eq:8}
\eeq
Using Equations \ref{eq:1}, \ref{eq:3} and \ref{eq:7}, the energy per
particle and the energy densities in regions (2) and (3) are given by
\beq
\ba{l}
{e_2 \over n_2} \simeq \sqrt{\Gamma_4 \over 2} {\omega_1^{3/4}
  \omega_4^{1/4} \over n_1}; ~~~{e_3 \over n_3} \simeq \sqrt{\Gamma_4 \over
  2} {\omega_1^{1/4} \omega_4^{3/4} \over n_4}; \\
e_2 = e_3 \simeq {\hat \gamma_2 \over \hat \gamma_2 -1} {\Gamma_4 \over 2}
\omega_1^{1/2} \omega_4^{1/2} = 2 \Gamma_4 \omega_1^{1/2}
\omega_4^{1/2},
\ea
\label{eq:9}
\eeq 
where in the last line we took $\hat \gamma_2 = \hat \gamma_3 = 4/3$. 

The results derived in Equation \ref{eq:9} can further be used to set
a minimum criteria on the Lorentz factor of the fast shell, $\Gamma_4$
that enables the existence of the two-shock system if the colliding
plasma shells are initially hot. Writing $e_i = \omega_i / \hat
\gamma_i + (\hat \gamma_i -1) n_i / \hat \gamma_i \approx \omega_i /
\hat \gamma_i$, the requirements $e_2 \geq e_1$ and $e_3 > e_4$ are
translated into the criteria
\beq
\left.
\ba{l}
\Gamma_4 \geq 2 {\hat \gamma_2 -1 \over \hat \gamma_1 \hat \gamma_2}
\sqrt{ \omega_1 \over \omega_4} = {3 \over 8} \sqrt{ \omega_1 \over
  \omega_4}, \\
\Gamma_4 \geq 2 {\hat \gamma_3 -1 \over \hat \gamma_4 \hat \gamma_3}
\sqrt{ \omega_4 \over \omega_1} = {3 \over 8} \sqrt{ \omega_4 \over
  \omega_1}, 
\ea
\right. \rightarrow \Gamma_4 \geq {3 \over 8} \max \left \{ \sqrt{ \omega_4 \over
  \omega_1}, \sqrt{ \omega_1 \over
  \omega_4} \right\},
\label{eq:10} 
\eeq
where we took $\hat \gamma_i = 4/3$ in all four regimes, which is
valid for hot plasmas. This minimum value of the Lorentz factor can be
understood as follows: as the Lorentz factor of the fast shell
($\Gamma_4$) decreases, eventually either the forward or reverse
shock ceases to be relativistic, and the amount of energy dissipated
from the (initially hot) plasma shells decreases. When this criterion
is met, (at least) one of the two shock waves ceases to
exist. Instead, a rarefaction wave will be created and propagate into
the hot plasma, while the second shock wave could still exist.

We emphasise that the result of Equation \ref{eq:10} sets the minimum
criteria on the Lorentz factor, originating from the physical
requirement that the shock waves are capable of dissipating the
shell's kinetic energy. To these criteria, one must add the underlying
assumption taken in this section, that $\Gamma_4 \gg \Gamma_2 \gg
1$. These criteria are further validated numerically (see
\S\ref{sec:3} below).

Finally, we note that a Lorentz transformation to the {\it lab frame},
in which $\Gamma_1^L \equiv \Gamma_1 \gg 1$, yields $\Gamma_2^{L} =
\Gamma_1 \Gamma_2 (1 +\beta_1 \beta_2) \simeq 2 \Gamma_1 \Gamma_2$,
and similarly $\Gamma_4^{L} \simeq 2 \Gamma_1 \Gamma_4$. The Lorentz
factor of the shocked plasmas are therefore given by (Equation
\ref{eq:8}),
\beq
\Gamma_2^L = \sqrt{\Gamma_1^L \Gamma_4^L} \left({\omega_4 \over
  \omega_1}\right)^{1/4}.
\label{eq:11}
\eeq

\subsection{Newtonian collision}
\label{sec:2.3}

We next consider the Newtonian (non-relativistic) case, in which the
relative motion between the two colliding shells is non- or
trans-relativistic at most. We do allow, though, the colliding plasmas
to be arbitrarily hot.  Similar to the relativistic treatment, we
initially assume $\Gamma_1 = 1$.


In this case, it is handy to define the internal energy (excl. rest
mass) $\epsilon_1$ by $\epsilon_1 \equiv e_1 - n_1$.  When writing the
energy density in region (1) as $e_1 = n_1 + \epsilon_1$, the forward
shock jump conditions (equations \ref{eq:1}, \ref{eq:2}) can be
written as a quadratic equation in the ratio of the proper densities
at both sides of the forward shock wave,
\beq
\ba{l}
\left({n_2 \over n_1} \right)^2 \left[ \Gamma_2 \left(1 + \hat
  \gamma_1 {\epsilon_1 \over n_1} \right) -1 \right]  \\
- \left({n_2
  \over n_1} \right) \left[ \left( 1 + \hat \gamma_1 { \epsilon_1
    \over n_1} \right) \left( 2 + {\hat \gamma_2 \over \hat \gamma_2
    -1} \Gamma_2^2 \beta_2^2 \right) - (\Gamma_2 +1) \right]  \\
+ \hat \gamma_1 \Gamma_2 {\epsilon_1 \over n_1}  =  0
\ea
\label{eq:12}
\eeq
 (with a similar equation holding for the reverse shock). Note that
Equation \ref{eq:12} is exact for all velocities. In order to obtain
useful approximation in the Newtonian regime, one needs to (i)
approximate $\Gamma_2 \simeq 1 + \beta_2^2/2$; and (ii) discriminate
between different regimes, based on the value of $\epsilon_1$: hot,
cool and cold.

In the {\bf hot regime}, $\epsilon_1/ n_1 \gg 1$. The ratio of densities
at both sides of the shock waves becomes
\beq
\left( {n_2 \over n_1} \right)_{\rm hot} \simeq 1 + {\beta_2 \over
  \sqrt{\hat \gamma_2 -1}} + {\beta_2^2 \over 2 (\hat \gamma_2 -1)} =
1 + \sqrt{3} \beta_2 + {3 \over 2} \beta_2^2
\label{eq:13}
\eeq
where we took $\hat \gamma_1 = \hat \gamma_2 \simeq 4/3$ in the last equality. Using
this result in Equation \ref{eq:1}, the energy density in region (2)
is given by
\beq
\ba{lcl}
e_2 & \simeq &  \epsilon_1 \left( 1 + {\hat \gamma_1 \over \sqrt{\hat
    \gamma_2 -1} } \beta_2 + {{\hat \gamma_1 \hat \gamma_2} \over 2 (\hat \gamma_2
  -1 )} \beta_2^2 \right) \\
&  = & \epsilon_1 \left( 1 + {4 \over \sqrt{3}}
\beta_2 + {8 \over 3} \beta_2^2 \right).
\ea
\label{eq:14}
\eeq
Clearly, a similar equation holds for the energy density in region (3). 

In the opposite, {\bf cool / cold} limit, namely $\epsilon_1/n_1 \ll
1$, we proceed as follows. First, in the non-relativistic limit,
$\beta_2 \ll 1$, the ratio of densities derived from the shock jump
conditions (Equation \ref{eq:12}) can be put in the form
\beq
\ba{l}
\left({n_2 \over n_1} \right)^2 \left[ {\beta_2^2 \over 2} + \hat
  \gamma_1 {\epsilon_1 \over n_1} \left(1 + {\beta_2^2 \over 2}
  \right) \right] \\ 
- \left({n_2 \over n_1} \right) \left[ \left( {\hat
    \gamma_2 + 1 \over \hat \gamma_2 -1} \right) {\beta_2^2 \over 2} +
  2 \hat \gamma_1 { \epsilon_1 \over n_1} \left( 1 + {\hat \gamma_2
    \over \hat \gamma_2 -1} {\beta_2^2 \over 2} \right) \right] \\
+ \hat \gamma_1 {\epsilon_1 \over n_1} \left( 1 + {\beta_2^2 \over 2} \right)
= 0.
\ea
\label{eq:15}
\eeq 

We next discriminate between the {\bf cool} case, in
which $\epsilon_1/ n_1 \gg \beta_2^2$, and the opposite, {\bf cold}
case, in which $\epsilon_1/ n_1 \ll \beta_2^2$.  In the {\bf cool}
scenario, the ratio of densities and the energy density in region (2)
are given by
\beq
\ba{l}
\left({n_2 \over n_1}\right)_{\rm cool} \simeq 1 + { 1 \over
  \sqrt{\hat \gamma_1 (\hat \gamma_2-1) {\epsilon_1 \over n_1}}}
\beta_2 + {3 - \hat \gamma_2 \over 4 \hat \gamma_1 (\hat \gamma_2
  -1){\epsilon_1 \over n_1}} \beta_2^2, \\
({e_2})_{\rm cool} \simeq n_1 \left[ 1 + {\epsilon_1 \over n_1} + {1 \over \sqrt{\hat
      \gamma_2 -1}}\left( \sqrt{\hat \gamma_1 {\epsilon_1 \over n_1}}
  +{1 \over \sqrt{\hat \gamma_1 {\epsilon_1 \over n_1}}} \right) \beta_2
  \right].
\ea
\label{eq:16}
\eeq
In this case, the value of the adiabatic index is not a-priori known
(see discussion in section \ref{sec:3} below).

In the {\bf cold} scenario, $\epsilon_1/n_1 \ll \beta_2^2$, similar
calculation yields
\beq
\ba{lcl}
\left({n_2 \over n_1}\right)_{\rm cold} & \simeq & \left( {\hat
  \gamma_2 + 1 \over \hat \gamma_2 -1} \right) + {2 \hat \gamma_1
  \over \hat \gamma_2 +1} {\epsilon_1 \over n_1} - {8 \hat \gamma_1
  \over (\hat \gamma_2 + 1)(\hat \gamma_2 -1)} {\epsilon_1 \over n_1
  \beta_2^2} \\
& = & 4 + \left( {5 \over 4} - {15 \over 2 \beta_2^2 }
\right) \left({\epsilon_1 \over n_1}\right), \\
({e_2})_{\rm cold} & \simeq & n_1 \left[ \left( {\hat \gamma_2 + 1
    \over \hat \gamma_2 -1} \right) \left(1 + {\beta_2^2 \over 2}
  \right) - {8 \hat \gamma_1 \over (\hat \gamma_2 + 1)(\hat \gamma_2
    -1) } \left({\epsilon_1 \over n_1 \beta_2^2}\right) \right. \\
 & & + \left. {3 \hat
    \gamma_1 + 1 \over \hat \gamma_2 + 1} \left({\epsilon_1 \over
    n_1}\right) \right] \\
 & = & n_1 \left[ 4 \left(1 + {\beta_2^2 \over 2} \right) + \left({9
    \over 4} - {15 \over 2 \beta_2^2} \right) \left({\epsilon_1 \over
    n_1 }\right) \right],
\ea
\label{eq:17}
\eeq
where we used $\hat \gamma_1 = \hat \gamma_2 = 5/3$ in the last
equality. We further point out that given an arbitrary value of $0 <
(\epsilon_1/n_1) < 1$, for very small relative velocities between the
plasma shells the plasma can be considered as ``cool'', while at
higher velocities it can be regarded as ``cold''.

The results presented in Equations \ref{eq:12} -- \ref{eq:17} are of
course symmetric with respect to the reverse shock, and are obtained
by replacing quantities in regions $(1), (2)$ with those in regions
$(4), (3)$, respectively, and exchanging $\beta_2$ with $\bar \beta_3
\simeq \beta_4 - \beta_2$. Using these replacements, one can use the
requirement $p_2 = p_3$ to determine the shocked plasma velocity,
$\beta_2 = \beta_3$ as a function of the colliding plasmas parameters
($n_1, e_1, n_4$ and $e_4$) as well as their relative velocities
$\beta_4$ in the different regimes. The results of the various
scenarios are summarized in Table \ref{tab:T1}.

\tabletypesize{\small}

\begin{deluxetable}{ll}
 \tablecolumns{2}
 \tablewidth{0pc}
 \tablecaption{Shocked plasma velocities in the various cases.}
\tablehead{ \colhead{Scenario }  & \colhead{Shocked shell velocities, $\beta_2 = \beta_3$}
}

\startdata 
(a)~cold $\rightarrow$ cold & $\beta_4 {\sqrt{n_4 \over
    n_1} \over \sqrt{n_4 \over n_1} + 1}$ \\ 
(b)~cold $\rightarrow$ cool & $ \beta_4 - \left[ {1 \over 2} \left(
  {\hat \gamma_2 -1 \over \hat \gamma_3 -1} \right) \left( {n_1 \over
    n_4} \right) \left({\epsilon_1 \over n_1} \right)\right]^{1/2}$ \\
(c)~cold $\rightarrow$ hot & same as (b), cold $\rightarrow$ cool \\
(d)~cool $\rightarrow$ cool & $ {(\hat \gamma_3 -1) \epsilon_4 - (\hat
  \gamma_2 -1) \epsilon_1 + (\hat \gamma_3 -1) \sqrt{ \hat \gamma_4
    \over \hat \gamma_4 -1} \sqrt{n_4 \epsilon_4} \beta_4 \over (\hat
  \gamma_2 -1) \sqrt{ \hat \gamma_1 \over \hat \gamma_1 -1} \sqrt{n_1
    \epsilon_1} + (\hat \gamma_3 -1) \sqrt{ \hat \gamma_4 \over \hat
    \gamma_4 -1} \sqrt{n_4 \epsilon_4}}$ \\
(e)~cool $\rightarrow$ hot & $ {(\hat \gamma_3 -1) \epsilon_4 - (\hat
  \gamma_2 -1) \epsilon_1 + (\hat \gamma_3 -1) \sqrt{ \hat \gamma_4
    \over \hat \gamma_4 -1} \sqrt{n_4 \epsilon_4} \beta_4 \over (\hat
  \gamma_2 -1) {4 \over \sqrt{3} \epsilon_1} + (\hat \gamma_3 -1)
  \sqrt{ \hat \gamma_4 \over \hat \gamma_4 -1} \sqrt{n_4 \epsilon_4}}$
\\
(f)~hot $\rightarrow$ hot & ${\epsilon_4 - \epsilon_1 + {4 \over
    \sqrt{3}} \epsilon_4 \beta_4 \over {4 \over \sqrt{3}} \left(
  \epsilon_1 + \epsilon_4 \right)}$ \\
\enddata
\label{tab:T1}
\end{deluxetable}

The results in Table \ref{tab:T1} can also be used to put constraints
on the minimum relative velocities between the shells ($\beta_4$) that
enables the formation of the double shock structure. These limits
originate from the requirements (a) $\beta_2 \geq 0$ and (b) $\bar
\beta_3 = \beta_4 - \beta_2 \geq 0$. In the first scenario considered
in Table \ref{tab:T1}, that of interaction between two cold plasmas,
$\beta_2$ is always smaller than $\beta_4$ and therefore there is no
restriction: a double shock structure will always form, for each value
of $\beta_4 > 0$. However, this is the exceptional case: in all other
scenarios, in which at least one of the shells is not completely cold,
such a restriction does exist. If $\beta_4$ is smaller then the
minimum value set by $n_1, n_4, \epsilon_1$ and $\epsilon_4$, the ram
pressure cannot compensate for the excess energy gained by
thermalization at the shock front. In these cases, two shocks cannot
form. Rather, similar to the relativistic case, a rarefaction wave
will form, which will gradually modify the properties of one of the
shells.

\section{Numerical Solution}
\label{sec:3}

In order to validate the analytical approximations presented in
Section \ref{sec:2} above as well as to investigate the intermediate
velocity (trans-relativistic) regime, we wrote a numerical code which
solves the dynamical conditions at each of the four regimes -
unshocked and shocked plasma shells that follow the collision of two
plasma shells.  The code simultaneously solves the set of twelve
coupled equations: three shock jump conditions each for the forward and
reverse shock waves, equating the pressure and velocity along the
contact discontinuity and four equations of state. The results are
obtained for a given set of six initial conditions: velocity, number
and energy densities in regions (1) and (4), the unshocked plasmas.

\subsection{Determination of the adiabatic indices in the different regimes}
\label{sec:3.1}

In order to account for the energy dependence of the adiabatic indices
$\hat \gamma_i$ in each of the four regimes, we use the prescription
derived by \citet{Service86}, which is accurate to $10^{-5}$.  Since
the classical gas law, $p_i = n_i T_i$ holds exactly in all regimes,
one can write
\beq
{e_i \over n_i} = T_i \left( {e_i + p_i \over p_i} - 1 \right).
\label{eq:3.1}
\eeq
We use the approximation derived by \citet{Service86},
\beq
\ba{lcl}
{p_i \over e_i + p_i} & = & 0.36 y + 0.036346 y^2 - 0.088763 y^3 \\
& & -0.047698 y^4 - 0.083547 y^5 + 0.073662 y^6,
\ea
\label{eq:3.2}
\eeq
where 
\beq
y \equiv {T_i \over 0.36 + T_i}. 
\label{eq:3.3}
\eeq
The results of Equation \ref{eq:3.2} are tabulated. Thus, for a given ratio
$e_i/n_i$, we use the tabulated results in Equation \ref{eq:3.1} to infer the
temperature $T_i$ in region $i$.

Once the temperature is known, at a second step the adiabatic index in
each regime is calculated using
\beq
\ba{lcl}
\hat \gamma_i & =  {1 \over 3} & \left( 5 - 1.21937 z + 0.18203 z^2 -
0.96583 z^3 \right. \\
 & & \left. + 2.32513 z^4 -2.39332 z^5+1.07136 z^6 \right)
\ea
\label{eq:3.4}
\eeq
where 
\beq
z \equiv {T_i \over 0.24 + T_i}. 
\label{eq:3.5}
\eeq

Calculation of the dynamical properties of the plasmas in the
different regimes is done as follows. We first guess a value of the
shocked plasma velocity (more precisely, of $\Gamma_2 \beta_2 =
\Gamma_3 \beta_3$), and solve for the two shock jump conditions. The
value of $\Gamma_2 \beta_2 $ is then varied, until the pressures at
each side of the contact discontinuity are equal.

In order to determine the adiabatic index in the shocked regions, for
each value of $\Gamma_2 \beta_2$ the shock jump conditions are solved
in iterative way. Following an initial guess of $\hat \gamma_2$, $\hat
\gamma_3$, the shock jump conditions are solved and the values of the
specific energies $e_2 / n_2$ and $e_3 / n_3$ are determined. The
values of the adiabatic index are then re-calculated, and the
calculation is repeated with the new value. We found that convergence
is typically very quick, within few iterations at most.

\subsection{Numerical results}
\label{sec:3.2} 

\begin{figure*}[ht!]
\centering
\begin{tabular}{c c}
 \includegraphics[width=0.45 \textwidth]{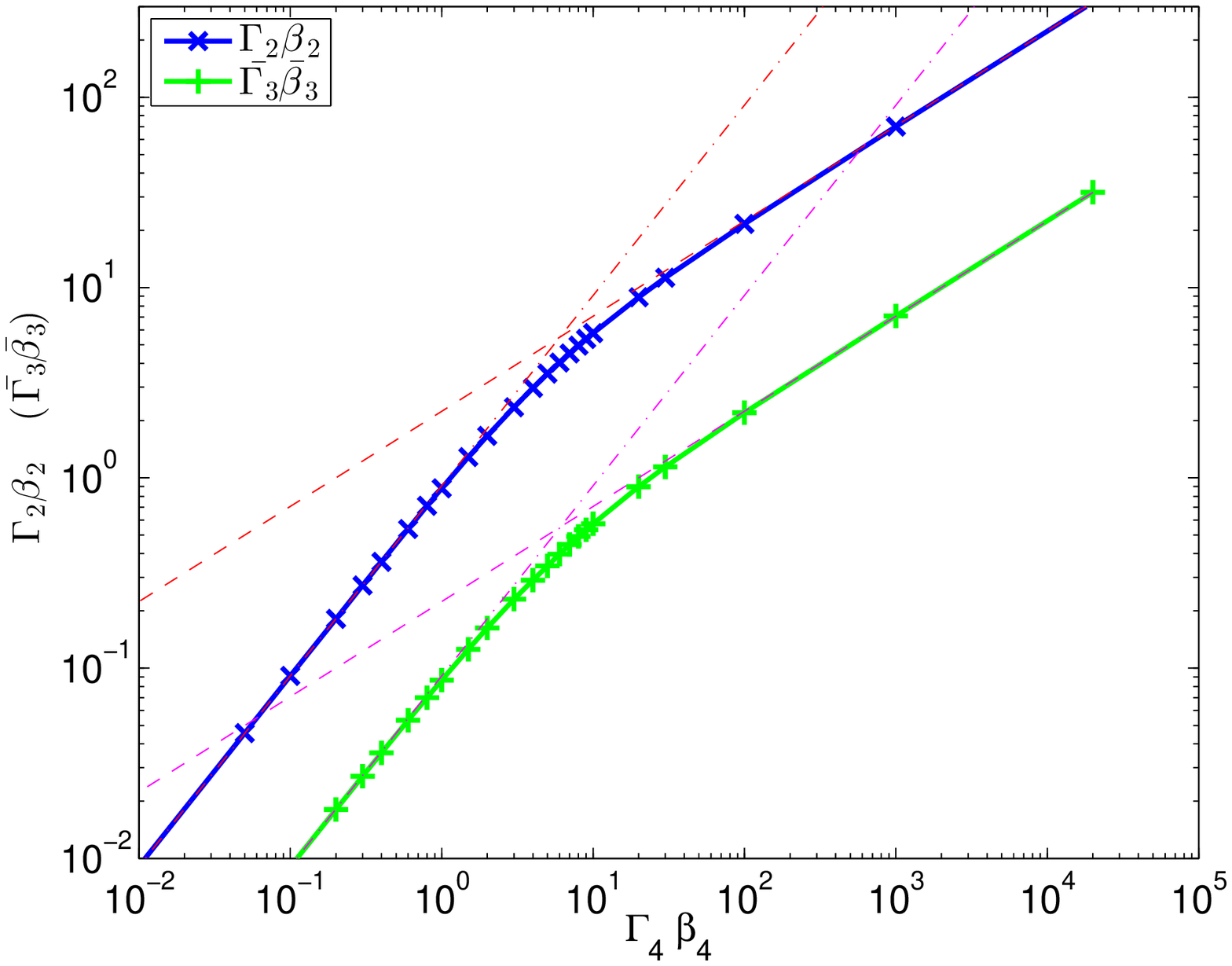} & 
 \includegraphics[width=0.45 \textwidth]{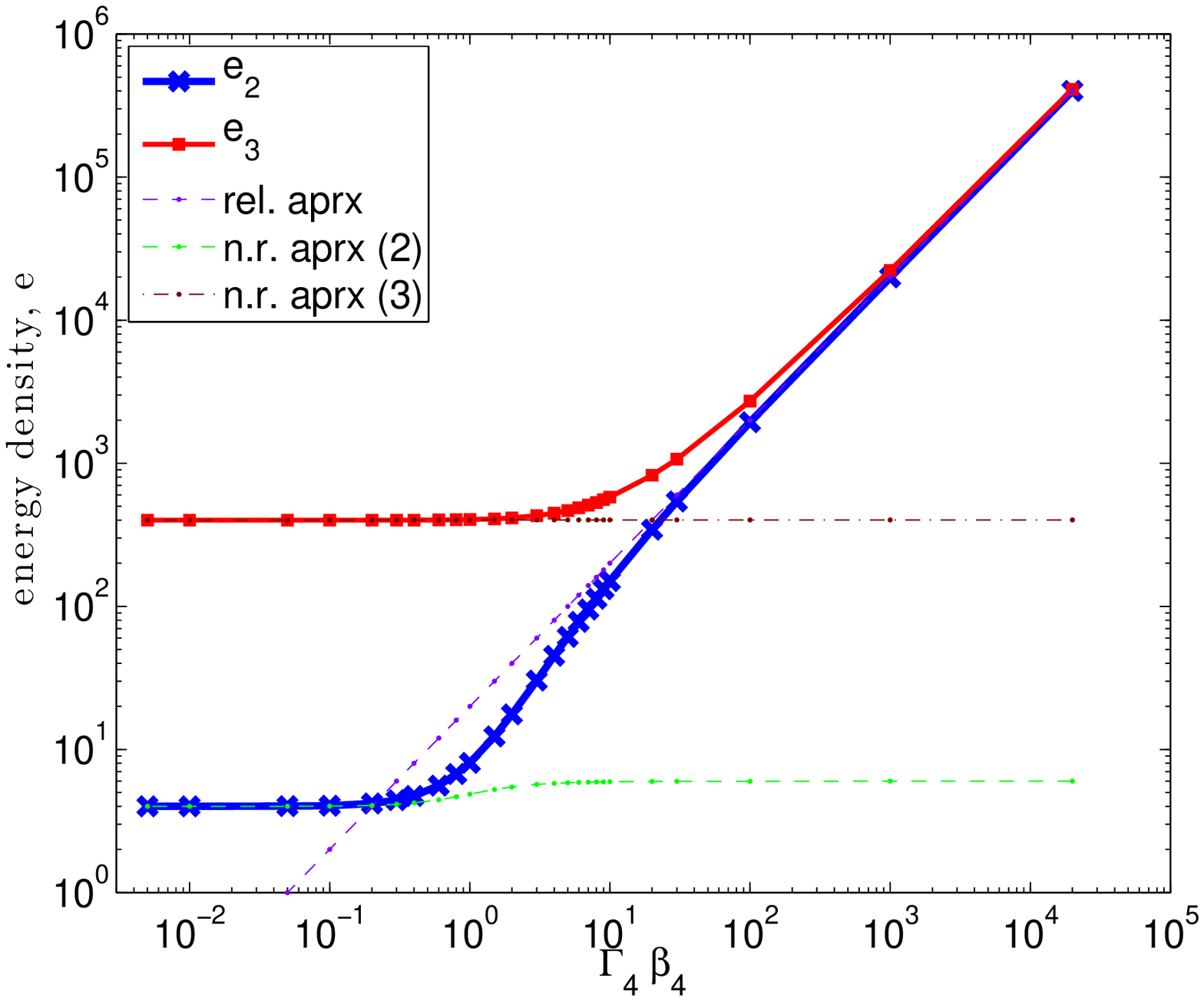}  \\
 \includegraphics[width=0.45 \textwidth]{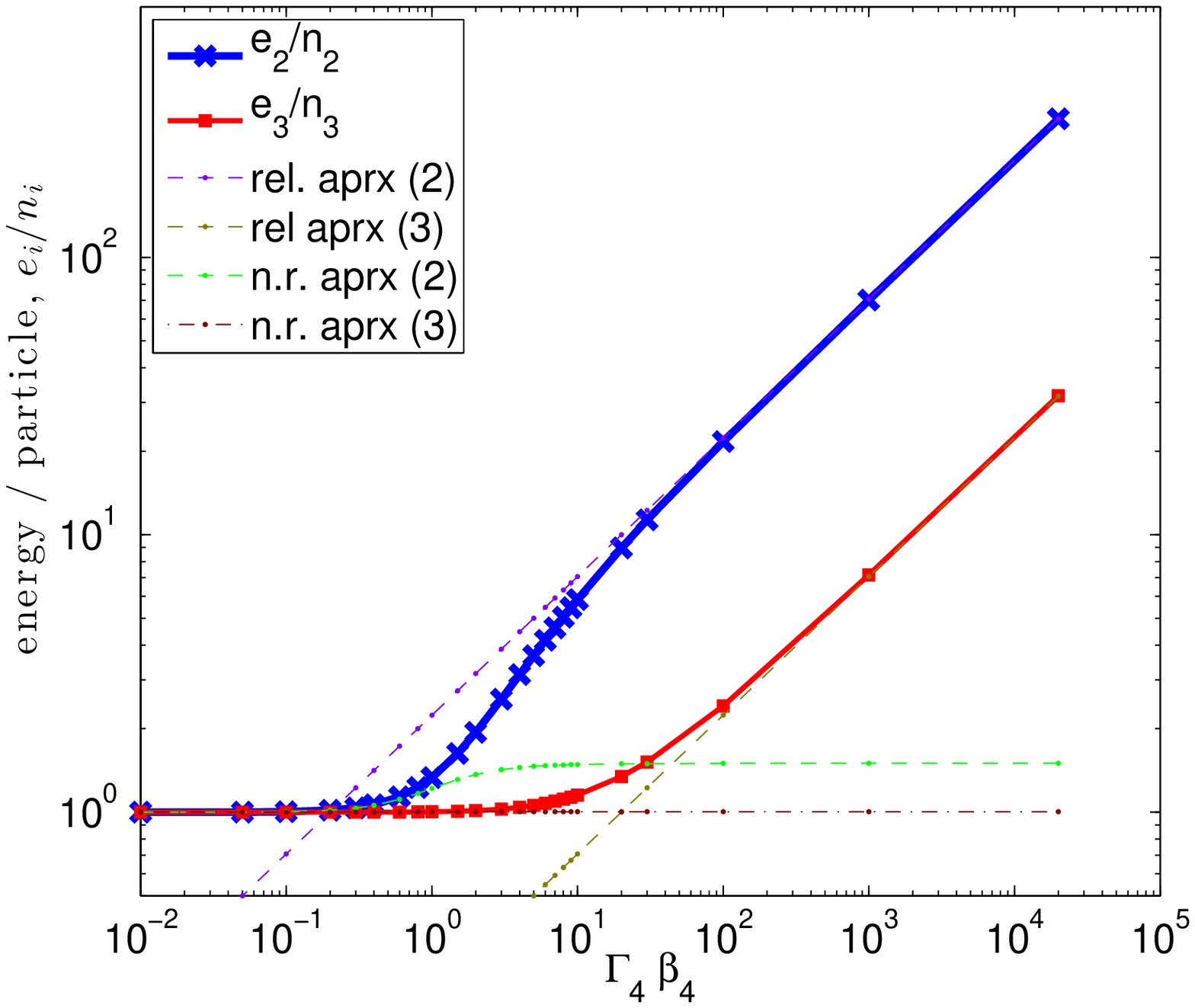} & 
\includegraphics[width=0.45 \textwidth]{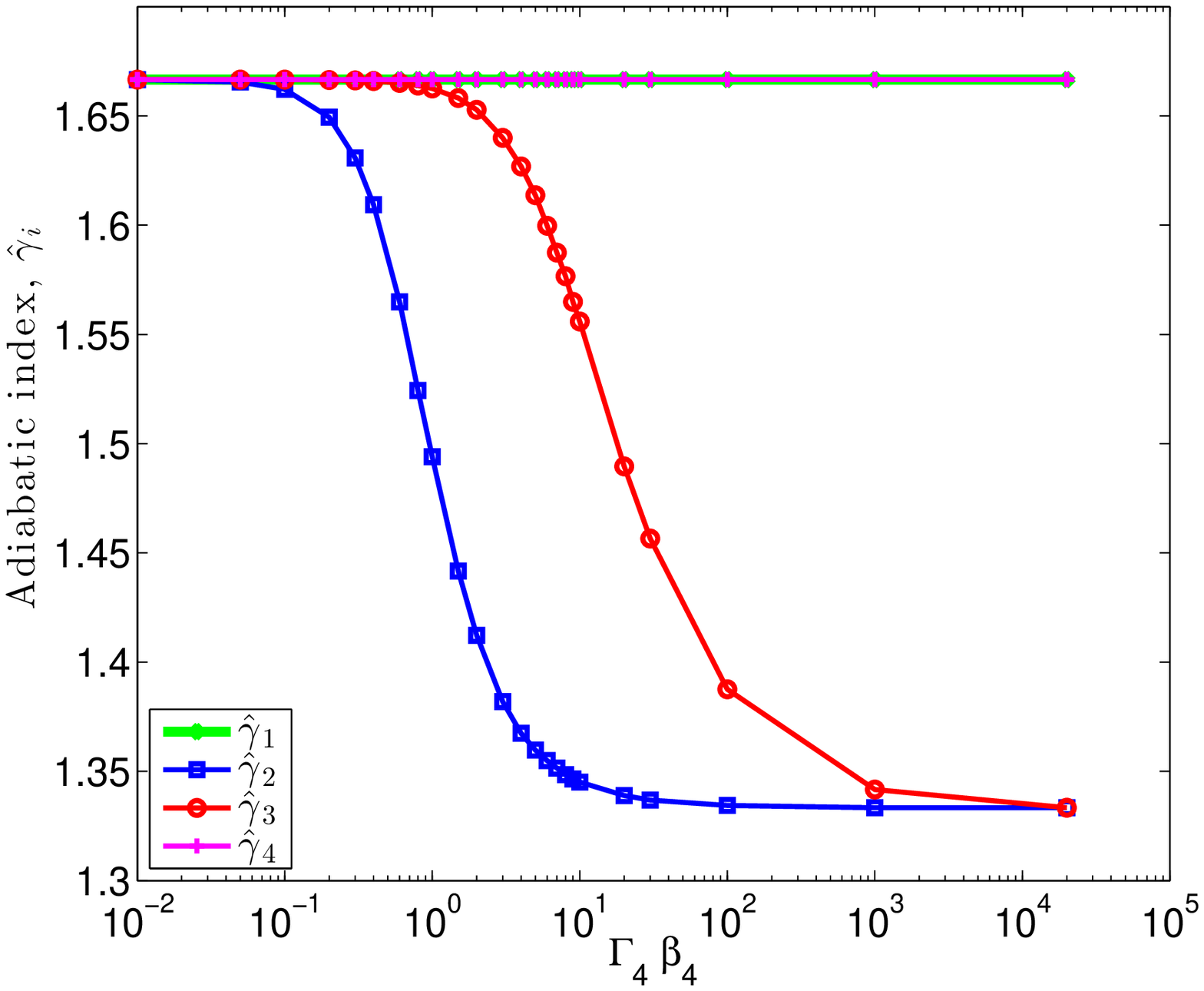}
\end{tabular}
  \caption{Velocities and thermodynamic properties of the shocked
    plasma following collision between two cold plasma shells
    ($\epsilon_i = 0$, $i=1,4$). Parameters considered are $n_1 = 1
    ~{\rm cm^{-3}}$, $\epsilon_1 = 0$, $n_4 = 100~{\rm cm^{-3}}$,
    $\epsilon_4 = 0$, $\beta_1 = 0$. Plasma parameters are shown as a
    function of the fast shell initial velocity, $\Gamma_4
    \beta_4$. Upper left: Velocities of the shocked plasma regions (2)
    and (3), as measured in the rest frame of slow plasma shell in
    region (1) ($\Gamma_2 \beta_2$) and the rest frame of the fast
    plasma shell in region (4), ($\bar \Gamma_3 \bar \beta_3$). Upper
    right: Energy densities in the shocked plasma regions (2) and
    (3). Lower left: Energy per particle, $e_i/n_i$ in the shocked
    plasma regions (2) and (3). Lower right: Adiabatic indeices $\hat
    \gamma_i$ in the four different regimes.}
    \label{fig:cold}
\end{figure*}

\begin{figure*}[ht!]
\centering
\begin{tabular}{c c}
 \includegraphics[width=0.45 \textwidth]{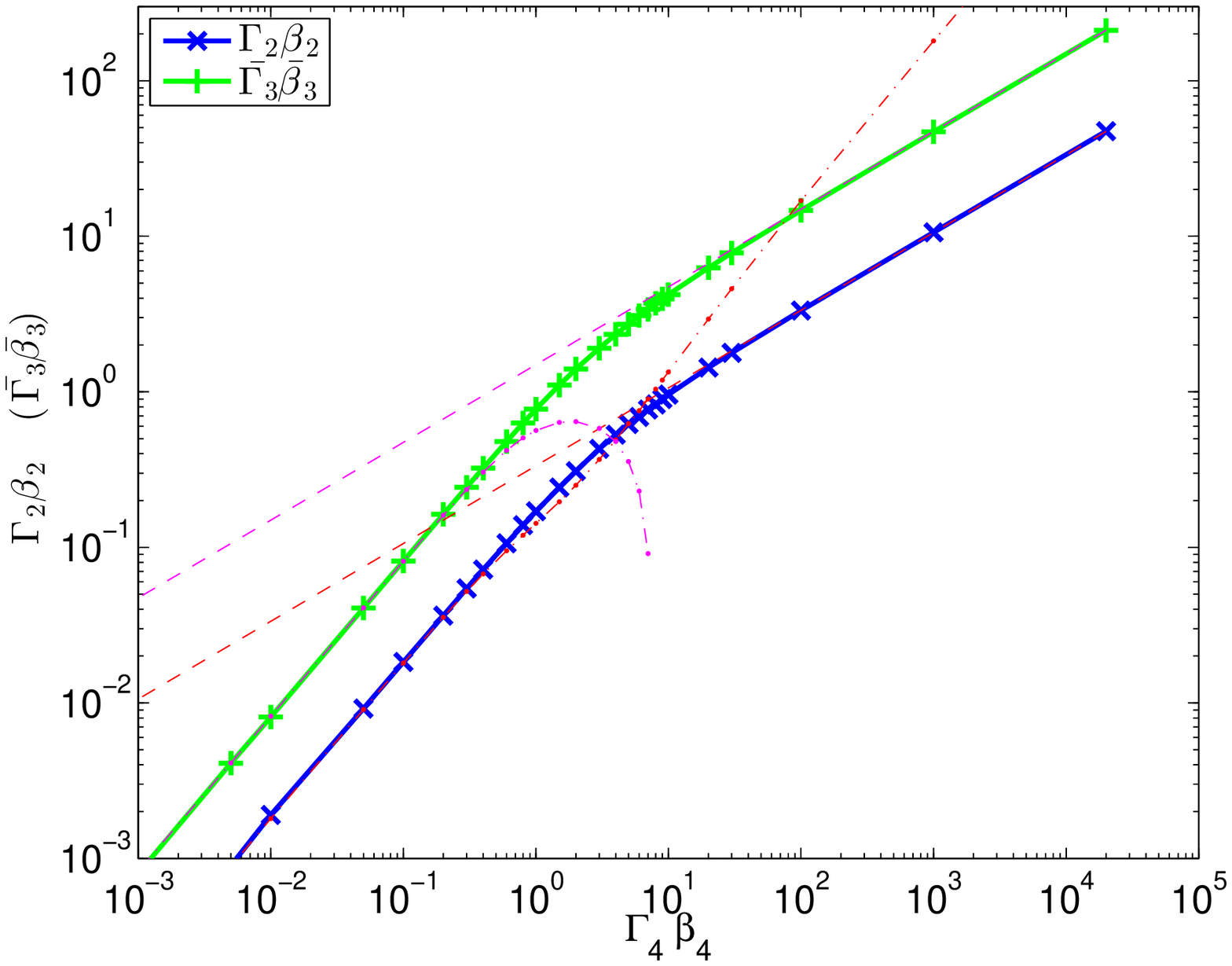} & 
 \includegraphics[width=0.45 \textwidth]{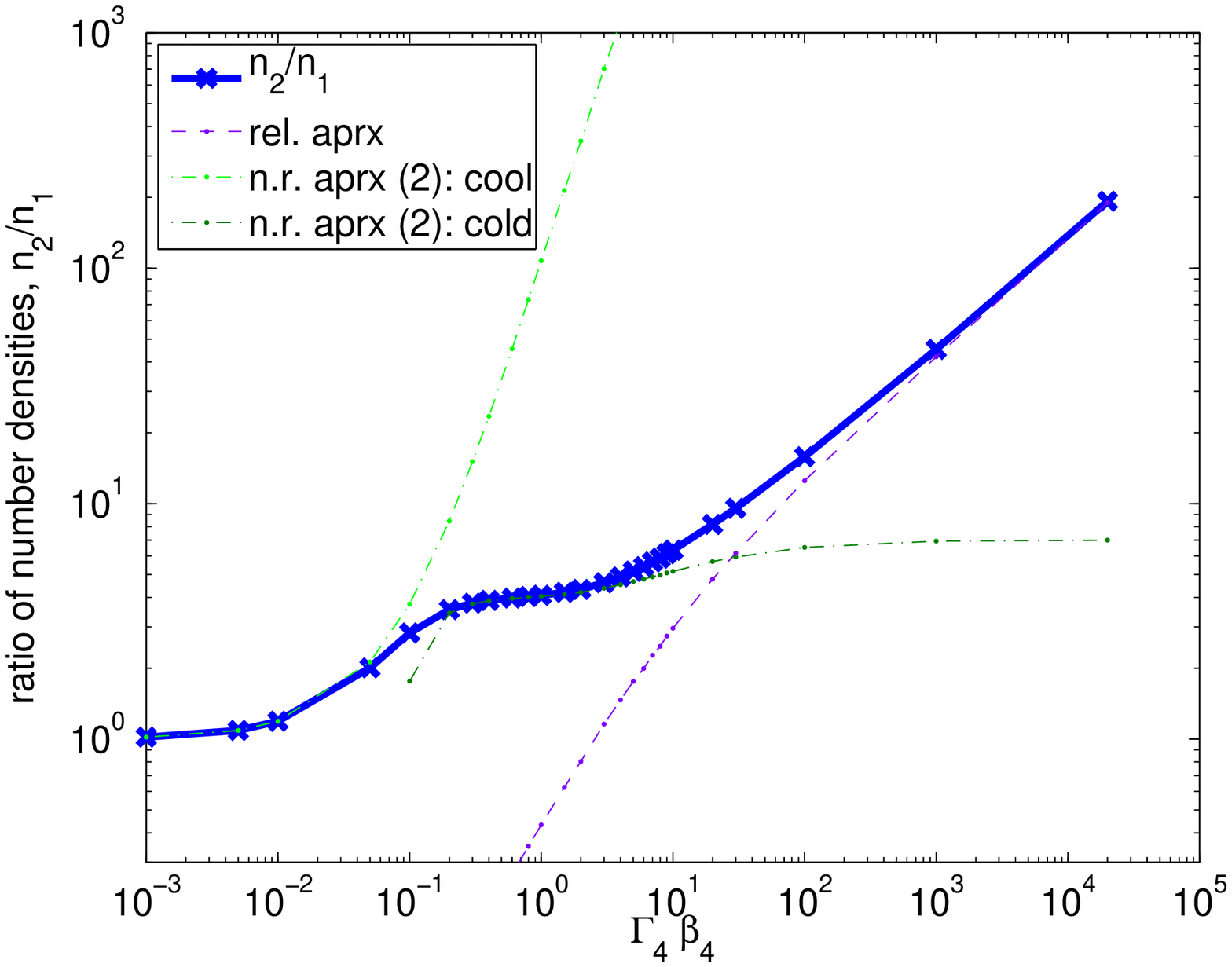}  \\
 \includegraphics[width=0.45 \textwidth]{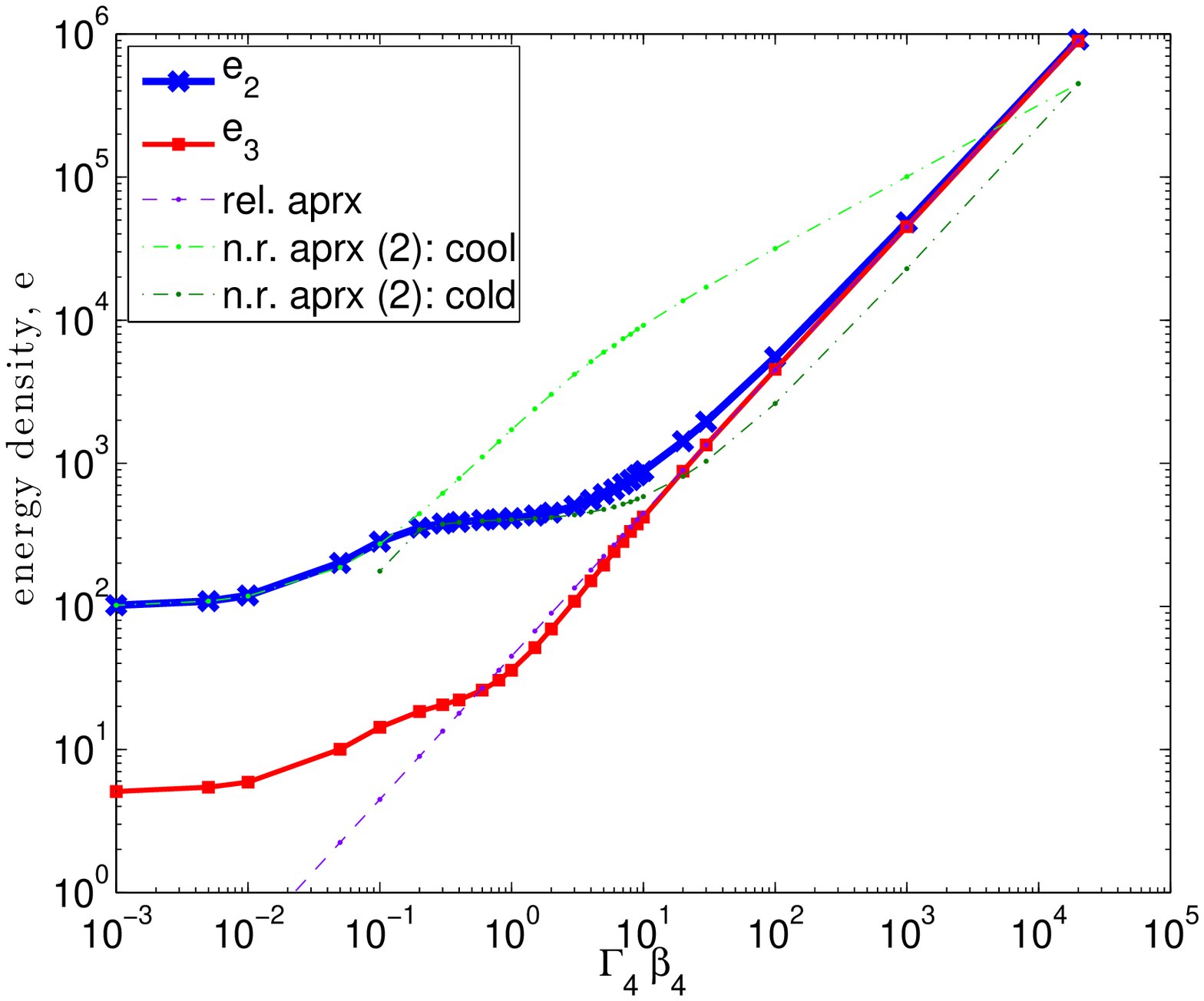} & 
\includegraphics[width=0.45 \textwidth]{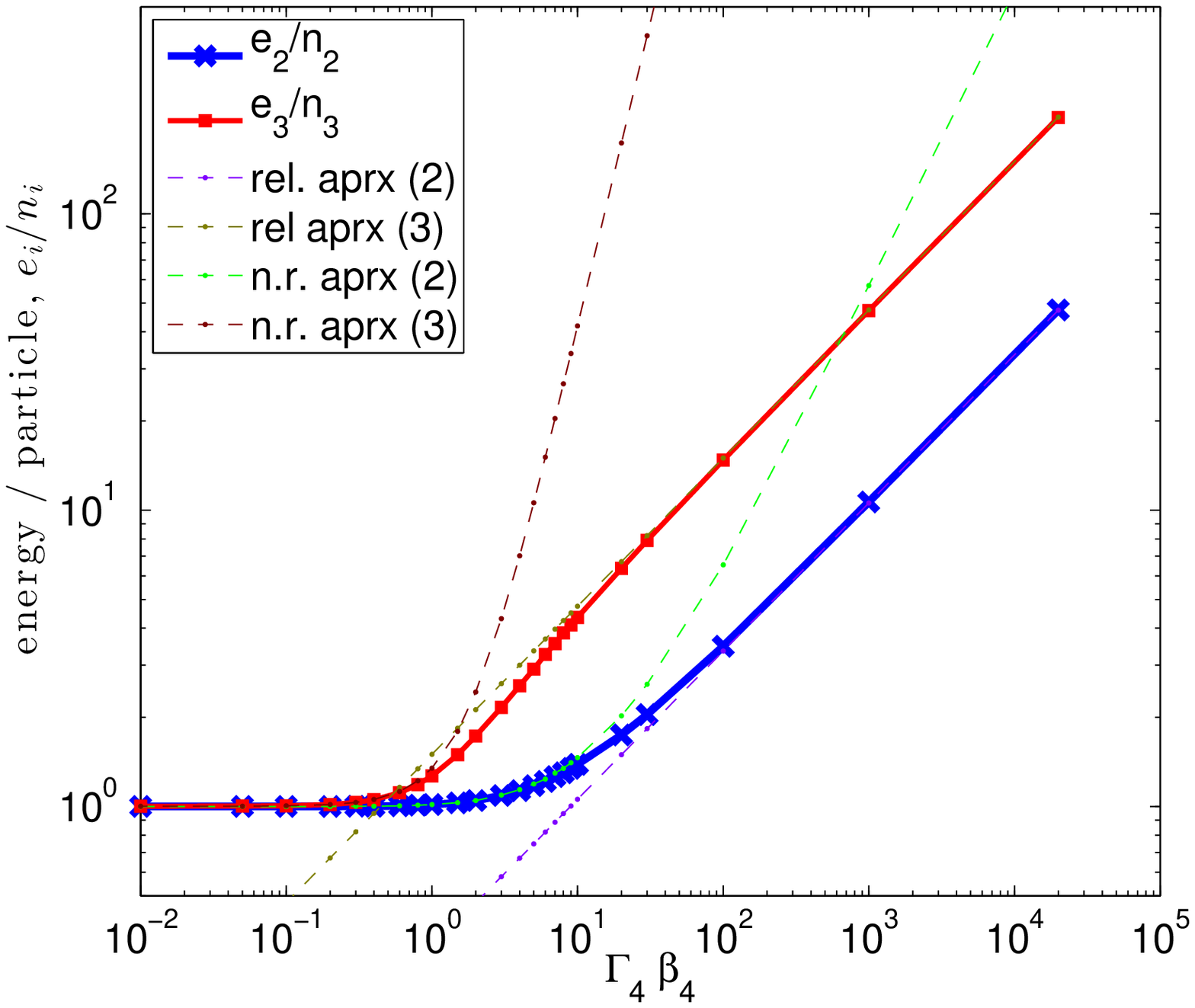}
\end{tabular}
 \caption{Velocities and thermodynamic properties of the shocked
   plasma following collision between two cool plasma shells
   ($\epsilon_i/n_i\ll 1$, $i=1,4$). Parameters considered are $n_1 =
   99.99 ~{\rm cm^{-3}}$, $\epsilon_1 = 0.01~{\rm erg~cm^{-3}}$, $n_4
   = 4.99~{\rm cm^{-3}}$, $\epsilon_4 = 0.01~{\rm erg~cm^{-3}}$,
   $\beta_1 = 0$. Plasma parameters are shown as a function of the
   fast shell initial velocity, $\Gamma_4 \beta_4$. Upper left:
   velocities of the shocked plasma (same as in Figure \ref{fig:cold}
   [upper left]), for collision of cool plasma shells. Upper right:
   Ratio of number densities across the forward shock, $n_2 /
   n_1$. Lower left: Energy densities in the shocked plasma regions
   (2) and (3). Lower right: Energy per particle, $e_i/n_i$ in the
   shocked plasma regions (2) and (3).}
    \label{fig:cool}
\end{figure*}

\begin{figure*}[ht!]
\centering
\begin{tabular}{c c}
 \includegraphics[width=0.45 \textwidth]{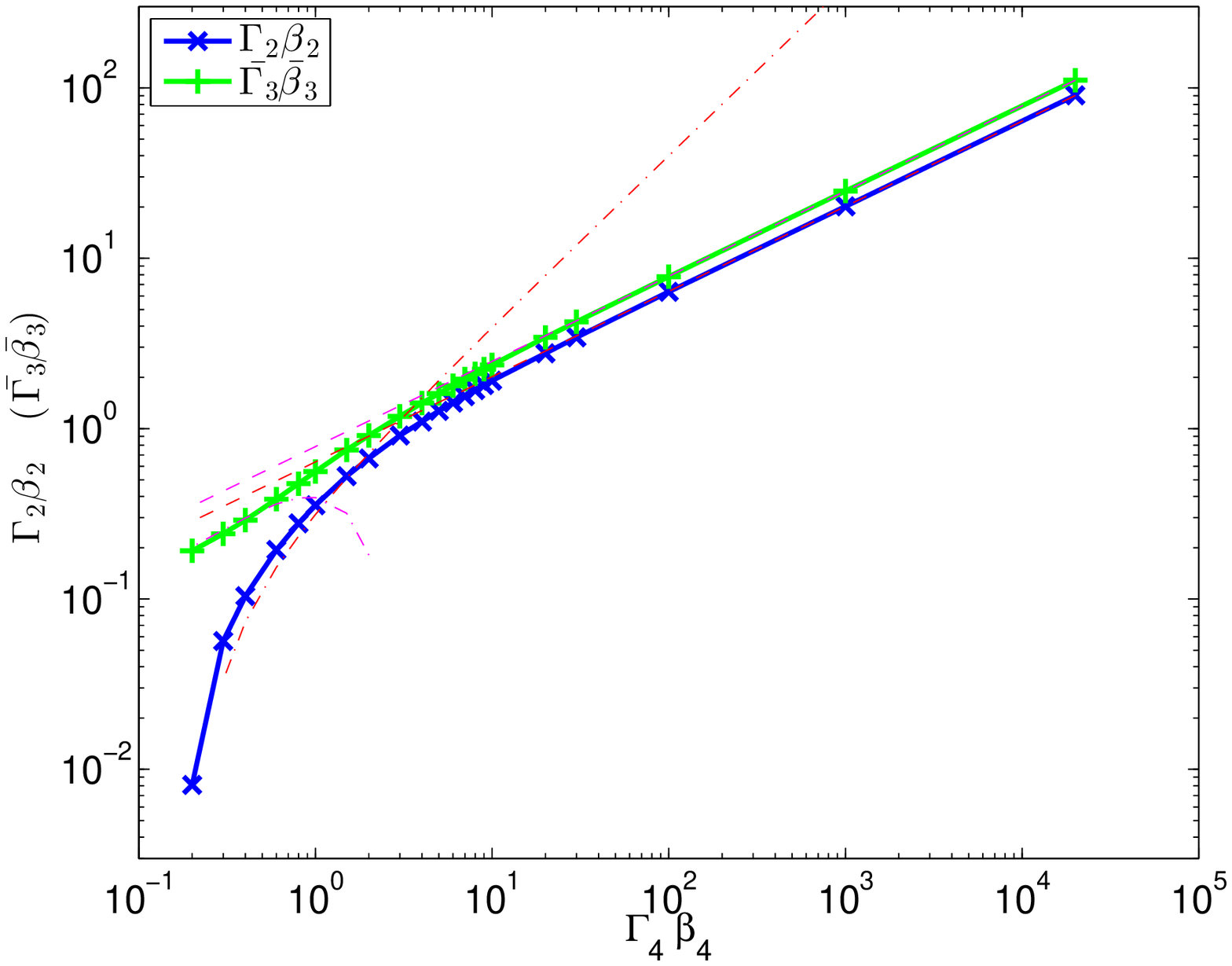} & 
 \includegraphics[width=0.45 \textwidth]{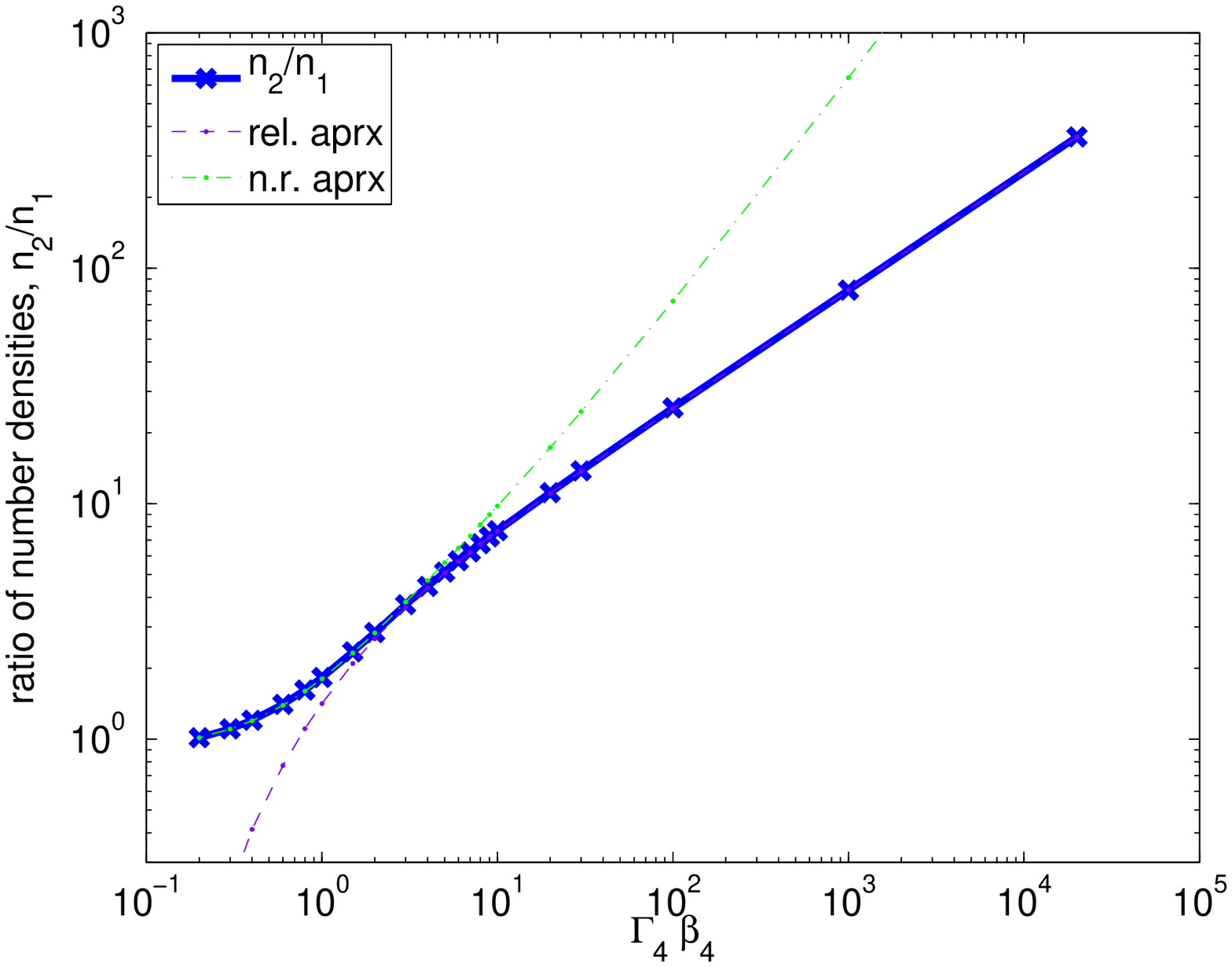}  \\
 \includegraphics[width=0.45 \textwidth]{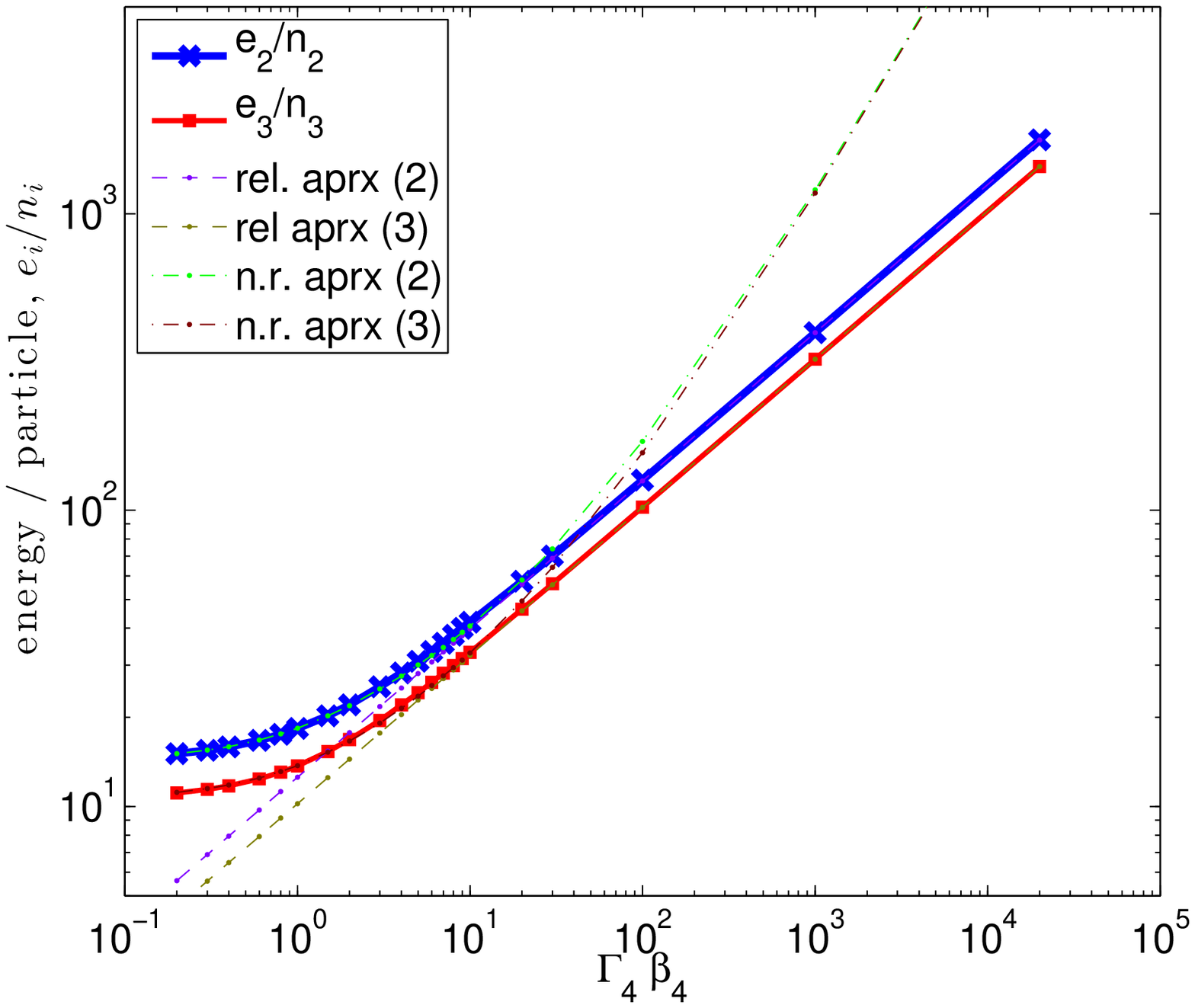} & 
\end{tabular}
 \caption{Velocities and thermodynamic properties of the shocked
      plasma following collision between two initially hot ($\epsilon_i/n_i \gg 1$) plasma
      shells. Parameters considered are $n_1 = 1 ~{\rm cm^{-3}}$,
      $e_1 = 15~{\rm erg~cm^{-3}}$, $n_4 = 1~{\rm cm^{-3}}$,
      $e_4 = 10~{\rm erg~cm^{-3}}$, $\beta_1 = 0$. Plasma
      parameters are shown as a function of the fast shell initial
      velocity, $\Gamma_4 \beta_4$. Upper left: Velocities of the shocked plasma (same as in Figure
          \ref{fig:cold} [upper left]), for collision of two hot plasma shells. Upper right: Ratio of number densities across the forward shock, $n_2 / n_1$. Lower left: Energy per particle, $e_i/n_i$ in the shocked plasma regions (2) and (3). }
    \label{fig:hot}
\end{figure*}

Examples of the numerical results, together with the analytical
approximations in the different regimes are presented in Figures
\ref{fig:cold} -- \ref{fig:hot}. 

In Figure \ref{fig:cold}, we calculate the dynamical and thermal
properties in all four regimes following the collision of two cold
plasma shells. The first (slow) shell is characterized by density $n =
1~{\rm cm^{-3}}$ and zero internal energy ($e_1 = n_1 m_e c^2$). The
fast plasma shell is characterized by higher density of $n_4 =
100~{\rm cm^{-3}}$ and is similarly cold, $e_4 = n_4 m_e c^2$. The
density contrast is chosen to be 100 for presentation purposes. We
further chose the slow plasma to be motionless, $\beta_1 =0$. The
results are presented as a function of $\Gamma_4 \beta_4$, where
$\beta_4$ is the relative velocity between the shells.

For cold plasmas as considered in Figure \ref{fig:cold}, there is no
lower limit on $\beta_4$, i.e., the two shock system always forms, for
any value of $\beta_4 > 0$. This system of cold plasmas is in fact
identical to the one considered already by \citet{SP95}. In the upper
left panel of Figure \ref{fig:cold} we show the shocked plasma
velocity, ($\Gamma_2 \beta_2$), in the rest frame of the slow shell as
well as the same velocity in the rest frame of the unshocked, fast
plasma in region (4), denoted by ($\bar \Gamma_3 \bar \beta_3$). The
asymptotic approximations in the relativistic (Equations \ref{eq:8})
and non-relativistic (Table \ref{tab:T1} (a)) regimes are given by the
dashed and dash-dotted lines. In producing the non-relativistic
approximation of the velocities, we replace $\beta_4$ with $\Gamma_4
\beta_4$. The results show excellent agreement - better than $\sim
10\%$ for $\Gamma_4 \beta_4 \leq 2$.

In upper right and lower left panels of Figure \ref{fig:cold} we show
the energy densities and the energy per particle ($e_i / n_i$) in the
shocked plasma regions (2) and (3) as a function of $\Gamma_4
\beta_4$. The analytic approximations in the relativistic regime
(Equation \ref{eq:9}) and non-relativistic regime (Equation
\ref{eq:17}) again provide an excellent description of the
thermodynamical properties of the plasma. The transition between the
non-relativistic and relativistic regimes occurs for those values of
$\Gamma_4 \beta_4$ in which $\Gamma_2 \beta_2 / \bar \Gamma_3 \bar
\beta_3$ becomes relativistic. Finally, in the lower right panel of
Figure \ref{fig:cold} we show the adiabatic indices in the different
regimes. While clearly $\hat \gamma_1 = \hat \gamma_4 = 5/3$, the
adiabatic indices of the shocked plasma are gradually changing as
$\Gamma_4 \beta_4$ increases, and the shocked plasma heats.

In Figure \ref{fig:cool} we consider a more complicated scenario, that
of a collision between two cool shells. We chose as parameters $n_1 =
99.99~{\rm cm^{-3}}$, $\epsilon_1 = 0.01~{\rm erg~cm^{-3}}$ (namely,
$e_1 = 100~{\rm erg~cm^{-3}}$), $n_4 = 4.99~{\rm cm^{-3}}$ and
$\epsilon_4 = 0.01~{\rm erg~cm^{-3}}$. Similar to the previous
example, we took $\beta_1 = 0$, namely slow shell at rest. These
values are chosen for presentation purposes, as we want to ensure a
good contrast of the shocked plasma properties between the different
regimes.

The velocities of the shocked plasma regions (2) and (3) as measured
in the rest frames of the slow plasma shell ($\Gamma_2 \beta_2$) and
the fast plasma shell ($\bar \Gamma_3 \bar \beta_3$) are shown in the
upper left panel of Figure \ref{fig:cool}. The analytical
approximation in the relativistic regime (Equation \ref{eq:8}) and the
non-relativistic regime (Table \ref{tab:T1} (d)) provide excellent
aproximations in the two regimes. The decay of the analytical
approximation to $\bar \beta_3 = \beta_4 - \beta_2$ around $\Gamma_4
\sim 1$ arises from the use of $\Gamma_4 \beta_4$ in the calculation
of $\beta_2$.

The ratio of densities across the forward shock is shown in the upper
right panel of Figure \ref{fig:cool}, together with the analytic
approximations. There are clearly three distinct regimes. First, there
is the relativistic regime, $\Gamma_2 \gg 1$. In this regime, the
density ratio is well approximated by the results given in Equation
\ref{eq:7}. A second regime is the non-relativistic, ``cold'' regime,
namely $\beta_2 \geq \sqrt{2 \epsilon_1/ n_1} = 0.014$ (in the
considered scenario), in which the density ratio is well approximated
by Equation \ref{eq:17}. Finally, when $\beta_2 \ll \sqrt{2 \epsilon_1
  /n_1}$, the approximation in the ``cool'' regime given in Equation
\ref{eq:16} provides a good fit to the density ratio. These same three
regimes are also clearly observed when considering the energy
densities of the shocked plasma in Figure \ref{fig:cool} (lower
left). Interestingly, when considering the ratio $e_i/n_i$ (Figure
\ref{fig:cool}, lower right) in the non-relativistic regime, the cool
and cold approximations can be combined to provide a good
approximation which reads $e_2/n_2 \simeq 1 + (\epsilon_1/n_1) +
(\Gamma_2 \beta_2)^2/2$. In the relativistic regime, this ratio is
well described by Equation \ref{eq:9}.

In Figure \ref{fig:hot} we provide a third example, that of a collision
between two initially hot plasma shells. As initial parameters, we
chose $n_1 =1~{\rm cm^{-3}}$, $e_1 = 15~{\rm erg~cm^{-3}}$, $n_4 =
1~{\rm cm^{-3}}$, $e_4 = 10~{\rm erg~cm^{-3}}$ and $\beta_1 = 0$.

For this choice of parameters, the results of Table \ref{tab:T1} (f)
show a minimum value of $\beta_4$, below which two shock waves cannot
form: For $\beta_4 = \sqrt{3} (\epsilon_1 - \epsilon_4)/4 \epsilon_4
\simeq 0.24$, $\beta_2 \rightarrow 0$. This is clearly demonstrated in
Figure \ref{fig:hot} [top left]. At larger relative velocity, the
results in Equation \ref{eq:8} and Table \ref{tab:T1} (f) provide an
excellent approximation to the shocked plasma velocity. The ratio of
number densities across the forward shock, $n_2/n_1$ (Figure
\ref{fig:hot}, top right) is well approximated by the analytical
approximations in Equation \ref{eq:7} (relativistic) and \ref{eq:13}
(non-relativistic). Similarly, the energy per particle in the shocked
regions (2) and (3) shown in the bottom left panel of Figure
\ref{fig:hot} are well approximated by the analytical result in
Equation \ref{eq:9} in the relativistic regime, and by $e_2/n_2 \simeq
(e_1/n_1) ( 1 + \Gamma_2 \beta_2 / \sqrt{3} + (\Gamma_2 \beta_2)^2/6)$
in the non-relativistic regime, which is readily derived from
Equations \ref{eq:13} and \ref{eq:14}.

\section{Observational Consequences}
\label{sec:4}

\subsection{Efficiency in kinetic energy dissipation}

The calculations above enable us to determine the efficiency of
kinetic energy conversion during two shell collisions.  Various
authors have calculated this effciency using an integral approach,
namely by determining the merged shell bulk Lorentz factor assuming a
plastic collision between the two shells, and using conservations of
energy and momentum.  For example, \citet{Kobayashi+97} and
\citet{Malzac14} considered cold plasma shells, while \citet{Spada+01}
and \citet{Jamil+10} generalized the result to hot plasmas.

Using the formulation developed here, we can generalize these
results. The calculation is done in the rest frame of the shocked
plasma. In this frame, the slow shell in region (1) is seen to have a
Lorentz factor $\tilde \Gamma_1$ (corresponding velocity $\tilde
\beta_1$), where $\tilde \Gamma_1 \tilde \beta_1 = \Gamma_1 \Gamma_2
(\beta_1 - \beta_2)$. The transfer of momentum from the slow shell
(region (1)) to the shocked plasma (region (2)), assuming a planar
symmetry along the $x$ direction is given by
\beq
dP^x(1) = \int dV'_1 T^{01} = \int dV'_1 \omega_1 {\tilde \Gamma_1}^2
\tilde \beta_1 = \int dV_1 \omega_1 \tilde \Gamma_1 \tilde \beta_1.
\label{eq:4.1}
\eeq
Here, $dV'_1 = dV_1 / \tilde \Gamma_1$ is the volume element of material in
region (1) that crosses the forward shock into region (2) per unit time,
as seen in the rest frame of shocked region (2), and $dV_1$ is the
same volume element as measured in the rest frame of region (1).

Using $\omega_1 = n_1 + \hat \gamma_1 \epsilon_1$, as well as $dM_1 =
\int dV_1 n_1$ and $dE_{th,1} = \int dV_1 \epsilon_1$, the momentum
transfer rate can be written as
\beq
dP^x(1) = \left(dM_1 + \hat \gamma_1 dE_{th,1}\right) \tilde \Gamma_1
\tilde \beta_1,
\label{eq:4.2}
\eeq
where we assumed that the velocities are not changed
during the shock propagation.

A similar calculation holds for the momentum transfer from the fast
plasma (region (4)), which could be written as
\beq
dP^x(4) = \left(dM_4 + \hat \gamma_4 dE_{th,4}\right) \bar \Gamma_3 \bar
\beta_3,  
\label{eq:4.3}
\eeq 
where $\bar \Gamma_3 \bar \beta_3 = \Gamma_4 \Gamma_2 (\beta_4 -
\beta_2)$. Equating the momentum transfer in both sides leads to the
velocity at the center of mass frame, which is the shocked fluid frame as
long as both shockes exist,
\beq
\beta_2 = {\Gamma_1 \beta_1 (dM_1 + \hat \gamma_1 dE_{th,1}) +
  \Gamma_4 \beta_4 (dM_4 + \hat \gamma_4 dE_{th,4}) \over \Gamma_1
  (dM_1 + \hat \gamma_1 dE_{th,1}) + \Gamma_4 (dM_4 + \hat \gamma_4
  dE_{th,4}) }.
\label{eq:26}
\eeq
In the ultra-relativistic case, $\Gamma_4 \gg \Gamma_2 \gg 1$ this can be written as
\beq
\Gamma_2 \simeq \left({ \Gamma_1 (dM_1 + \hat \gamma_1 dE_{th,1}) +
  \Gamma_4 (dM_4 + \hat \gamma_4 dE_{th,4}) \over (dM_1 + \hat
  \gamma_1 dE_{th,1}) /\Gamma_1 + (dM_4 + \hat \gamma_4
  dE_{th,4})/\Gamma_4} \right)^{1/2}.
\label{eq:4.4}
\eeq
This result differs from the result that appears in
\citet{Spada+01} (their Equation (4)) as well as in \citet{Jamil+10},
by the inclusion of the adiabatic indices $\hat \gamma_i$ that
multiply the thermal energies, which are omitted in these
works. These can be traced back to the inclusion of the pressure term
in the shocked plasma.

We further point out that equating the momenta transfer from regions
(1) and (4) using equations \ref{eq:4.2} and \ref{eq:4.3} in the
relativistic case, would retrieve back Equation \ref{eq:8}. These
results imply that the efficiency of kinetic energy conversion as
calculated in \citet{Spada+01} and \citet{Jamil+10} hold, provided
that the final Lorentz factor is calculated using Equation
\ref{eq:4.4}.

\subsection{Basic scalings of synchrotron emission}

The heated shocked plasma will radiate its energy. The observed signal
can therefore be used as a probe of the initial, unshocked plasma
shells properties. Full radiative calculations require additional
parameters, such as the exact value of the magnetic field as well as
assumptions about the radiating paricles distribution in the shocked
plasma regions, and are therefore left for a future work.

Here, we provide some basic scaling laws of the characteristic
frequencies expected from synchrotron emission, which is likely the
easiest (and most commonly discussed) signal that can be detected, and
can therefore be used to probe the plasma conditions. These are
particularly simple in the relativistic regime, where the plasma is
substantially heated by the shock waves. We therefore focus here in
the relativistic regime.

We scale the properties of the synchrotron emission in region $i$ by
adopting the common assumption that magnetic fields are generated by
the shock waves, and that the generated magnetic energy density is
some constant fraction of the post-shock thermal energy density,
namely $B_i^2 \propto e_i$. Furthermore, we assume that the electrons
carry some constant fraction of the proton energy, resulting in
electron's Lorentz factor $\gamma_{el,i} \propto (e_i/n_i)$. As the
characteristic synchrotron emission frequency is $\nu_{syn,i} \propto
B_i \gamma_{el,i}^2$, one finds the scaling
\beq
{\nu_{syn,2} \over \nu_{syn,3}} \propto {e_2^{1/2} \left({e_2 \over
    n_2}\right)^{2} \over e_3^{1/2} \left({e_3 \over n_3}\right)^{2}}
= \left({n_4 \over n_1} \right)^2 \left({\omega_1 \over \omega_4}
\right).
\label{eq:4.5}
\eeq 

If we denote by $\Delta_1$ and $\Delta_4$ the (comoving) widths of the
colliding shells, the total number of radiating electrons is $N_1
\propto n_1 \Delta_1$ and $N_4 \propto n_4 \Delta_4$ (under the 1-d
assumption). Since the total observed power is $P_{syn} \propto N B^2
\gamma_{el}^2$ \citep{RL79}, the ratio of synchrotron power between
the two shocked regions is therefore
\beq
{P_{syn,2} \over P_{syn,3}} = { n_1 \Delta_1 e_2 \left({e_2 \over
    n_2}\right)^{2} \over n_4 \Delta_4 e_3 \left({e_3 \over
    n_3}\right)^{2} } = \left({\Delta_1 \over \Delta_4}\right)
\left({n_4 \over n_1} \right) \left({\omega_1 \over \omega_4}\right).
\label{eq:4.6}
\eeq 
 
In the relativistic scenario, the observed time scale for the forward
shock wave to cross the slow plasma shell is $ \sim \Delta_1 \Gamma_1
/c$, while the time scale of the reverse shock to cross the fast
plasma is $\approx \Delta_4 \Gamma_2^2/\Gamma_4 c \sim 2 \Gamma_1
(\omega_4/\omega_1) \Delta_4 c$ \citep[e.g.,][]{SP95}. Thus, the
observed ratio of the time scale of existence of the two shock waves
is
\beq
{t_{fs} \over t_{rs}} = \left({\Delta_1 \over \Delta_4}\right)
\left({\omega_1 \over \omega_4}\right).
\label{eq:4.7}
\eeq

These results imply that identification of the ratios of the three
main characteristics of synchrotron emission from the forward and
reverse shock waves, namely the characteristic frequency, total power
and time scales, are sufficient to provide direct information about
the ratio of number densities, enthalpies and initial sizes of the
colliding shells. Interestingly, in the ultra-relativistic limit,
these results are independent on the unknown Lorentz factor. As we
showed above, using these initial conditions one can calculate the
properties of the merged shell. Therefore, direct observations of
multiple shell collisions could provide information about two key
ingredients. The first is the initial conditions of the ejected
shells, hence the properties of the inner engine. The second is the
temporal, hence spatial evolution (adiabatic losses) of the merged
shell.

\section{Summary and discussion}
\label{sec:5}

In this work, we considered the collision of two plasma shells, as is
expected in the ``internal shock'' model.  We generalized previous
treatments of the problem by considering plasmas which can be
arbitrarily hot. This is a natural consequence of the internal shocks
scenario, as, after the first collision, the merged shell is
inevitably hot (and can be very hot if the shells are relativistic,
see Equation \ref{eq:9}). We point out that while in between
collisions the colliding shells lose their energy adiabatically, the
decrease in temperature (or energy per particle) is $(e/n) \propto T
\propto r^{-2/3}$, and thus even if the internal collisions occur
within a range of several orders of magnitude in radii, adiabatic
cooling is not sufficient to completely cool the plasma shells.

We derived analytical approximations for the shocked shell velocities
in both the relativistic (Equation \ref{eq:8}) and non-relativistic
(Table \ref{tab:T1}) regimes. A very important result we found is that in the
general scenario (as opposed to the cold scenario) there is a minimum
relative velocity, or Lorentz factor, that enables the formation of the
two shock system (Equation \ref{eq:10}). The physical reason for this
is the requirement of the ram pressure to exceed the pressure
associated with the excess of thermal energy caused by the shock. If
this criterion is not met, only a single shock wave is expected, while
a rarefaction wave will propagate into the hotter plasma. In this
case, we expect the radiative signal to be much weaker.

We further provided analytical expressions for the energy density and
for the energy per particle in the shocked region. We found that for
non-relativistic collision, one needs to discriminate between three
scenarios: ``hot'' plasma, for which $\epsilon/n \gg 1$, ``cool''
plasmas for which $1 \gg \epsilon/n \gg \beta^2$, and ``cold'' plasma,
for which $1 \gg \beta^2 \gg \epsilon/n$. We provided the analytical
expressions for  thermodynamical properties of the shocked plasma in each
of these cases.

We discussed several observational consequences of the dynamical
results. We showed that in calculating the final Lorentz factor of the
merged shell, hence the efficiency of kinetic energy dissipation, one
needs to consider the pressure of the shocked plasma. We provided the
basic scaling laws of synchrotron emission in the ultra-relativistic
regime, and showed that measurements of the peak energy, flux and time
scale of emission enables one to deduce important information about
the initial shells properties, as well as the spatial evolution of the
propagating shells.

The results provided here emphasise the fact that the properties of
the shocked plasma depend not only on the relative velocities between
the colliding plasma shells, but also on the energy per particle in
each colliding shell. These results are therefore important in the
study of signals from multiple collisions that are expected in various
environments, such as GRBs, XRBs and AGNs. Furthermore, our numerical
results are particularly useful for probing the plasma properties in
the trans-relativistic regime, which is likely the dominant regime in
XRBs and possibly AGNs. As we demonstrated in \S \ref{sec:3} above,
while no simple analytical expressions exist in this regime, still
reasonable analytical fits do exist, and can be very useful in
understanding the underlying properties of these objects.

The results obtained in this work imply that the overall efficiency of
kinetic energy dissipation in a multiple shock scenario is in general
different than previous calculations that considered collisions
between cold shells \citep{Kobayashi+97, DM98, LGC99, Kumar00,
  Beloborodov00, GSW01, Ioka+06}. In a realistic scenario of hot
shells, when estimating the efficiency in multiple shells collisions,
one needs to consider: (i) the properties of each shell immediately
after the collision; (ii) the adiabatic cooling of the shells in
between the collisions; and (iii) adiabatic expansion of each shell in
between collisions, which results in differential velocity field
within the expanding shell \citep{KS01}. In this work we focus on part
(i) of this problem. We leave a complete treatment of a multiple shell
collision scenario for a future work.

\acknowledgments AP wishes to thank Damien B\'egu\'e, Felix Ryde, Ralph
Wijers and Bing Zhang for useful comments. This research was partially
supported by the European Union Seventh Framework Programme
(FP7/2007-2013) under grants agreement ${\rm n}^\circ$ 618499 (AP) and ${\rm n}^\circ$ 322259 (PC).

\bibliographystyle{/Users/apeer/Documents/Bib/apj}


\end{document}